# A Cs-Based Optical Frequency Measurement Using Cross-Linked Optical and Microwave Oscillators

Chr. Tamm, N. Huntemann, B. Lipphardt, V. Gerginov, N. Nemitz[*], M. Kazda, S. Weyers, E. Peik

*Physikalisch-Technische Bundesanstalt, Bundesallee 100, 38116 Braunschweig, Germany*

**ABSTRACT**

We describe a measurement of the frequency of the $^2S_{1/2}(F = 0) - {}^2D_{3/2}(F' = 2)$ transition of $^{171}$Yb$^+$ at the wavelength 436 nm (frequency 688 THz), using a single Yb$^+$ ion confined in a Paul trap and two caesium fountains as references. In one of the fountains, the frequency of the microwave oscillator that interrogates the caesium atoms is stabilized by the laser that excites the Yb$^+$ reference transition with a linewidth in the hertz range. The stability is transferred to the microwave oscillator with the use of a fiber laser based optical frequency comb generator that also provides the frequency conversion for the absolute frequency measurement. The frequency comb generator is configured as a transfer oscillator so that fluctuations of the pulse repetition rate and of the carrier offset frequency do not degrade the stability of the frequency conversion. The phase noise level of the generated ultrastable microwave signal is comparable to that of a cryogenic sapphire oscillator. For fountain operation with optical molasses loaded from a laser cooled atomic beam source, we obtain a stability corresponding to a fractional Allan deviation of $4.1 \cdot 10^{-14} (\tau/\mathrm{s})^{-1/2}$. With the molasses loaded from thermal vapor and an averaging time of 65 h, we measure the frequency of the Yb$^+$ transition with a relative statistical uncertainty of $2.8 \cdot 10^{-16}$ and a systematic uncertainty of $5.9 \cdot 10^{-16}$. The frequency was also simultaneously measured with the second fountain that uses a quartz-based interrogation oscillator. The unperturbed frequency of the Yb$^+$ transition is realized with an uncertainty of $1.1 \cdot 10^{-16}$ that mainly results from the uncertainty of the blackbody shift at the operating temperature near 300 K. The transition frequency of 688 358 979 309 307.82(36) Hz, measured with the two fountains, is in good agreement with previous results.

[*]Present address: Quantum Metrology Laboratory, RIKEN, Wako, Saitama, Japan.

# I. INTRODUCTION

Currently the most precise measurements carried out in the framework of the international system of units (SI) are measurements of the absolute frequencies of optical frequency standards by means of a caesium fountain clock. Here the Cs clock provides a reference frequency in the radiofrequency (rf) range that is directly linked to the ground-state hyperfine interval of $^{133}$Cs, $\nu_{Cs} \approx 9.2$ GHz, which defines the SI second. Several Cs fountains reach a relative frequency uncertainty in the $10^{-16}$ range [1]. Some optical standards based on reference transitions of single ions in radiofrequency traps or neutral atoms in an optical lattice have considerably lower uncertainties, the most advanced reaching the $10^{-17}...10^{-18}$ range [2-10]. Measurements of the frequency ratios between these optical clocks and of their absolute frequencies prepare the ground for a future redefinition of the SI second on the basis of an optical reference transition. The frequency conversions that are required in such measurements are usually realized with the use of an optical frequency comb generator, a laser source with a broad equidistant mode spectrum that can be unambiguously mapped to two frequencies in the rf range [11,12].

In absolute frequency measurements of optical clocks, the averaging time needed to reach a targeted statistical uncertainty is determined by the instability of the fountain reference because it is the dominant noise source. Generally the stability performance of frequency standards that rely on the cyclic interrogation of an ensemble of quantum absorbers can be estimated through the fractional Allan deviation $\sigma_y(\tau)$ as

$$\sigma_y(\tau) \approx \frac{\delta\nu}{\nu} \cdot \frac{1}{\pi} \cdot \frac{1}{\text{SNR}} \cdot \sqrt{\frac{T_C}{\tau}} \ . \tag{1}$$

Here $\tau$ is the averaging time, $T_C$ the duration of one measurement cycle, $\delta\nu/\nu$ the normalized linewidth of the atomic resonance signal, and SNR the signal-to-noise ratio obtained in a single measurement. With optical clocks that use ultrastable interrogation lasers, it is possible to obtain $\delta\nu/\nu < 10^{-14}$ while for Cs fountains the lower frequency and the limited interaction time lead to $\delta\nu/\nu_{Cs} < 10^{-10}$, resulting in a significantly lower stability. The stability of fountain clocks can be enhanced by probing larger ensembles of Cs atoms under conditions where the signal-to-noise ratio of the atomic resonance signal is close to the limit given by quantum projection noise [13]. Here one of the most demanding tasks is to generate a microwave interrogation signal whose spectral purity is sufficient to avoid excess noise [14]. The only kind of microwave oscillator that can provide the required low intrinsic phase noise level is the cryogenic sapphire oscillator [15-17]. One such system has been successfully operated for more than 10 years and a stability of $\sigma_y(\tau) = 4.1 \cdot 10^{-14} (\tau/\text{s})^{-1/2}$ has been achieved in a recent absolute frequency measurement of optical lattice clocks [9,18]. Alternatively, microwave signals with very low phase noise can also be generated by combining an optical frequency comb generator (OFC) with an ultrastable laser [19-22]. It has been shown that the stability of Cs fountains that use OFC-stabilized interrogation oscillators can be comparable to that obtained with a cryogenic sapphire oscillator [20,23,24].

In this paper we demonstrate a new scheme for absolute optical frequency measurements where an OFC is configured as a transfer oscillator that links the optical and microwave interrogation oscillators and serves two functions at the same time: it stabilizes the microwave oscillator and it yields the relation between the optical frequency and $\nu_{Cs}$ as determined by the fountain's resonance signal. The transfer oscillator scheme relies on combining an OFC with a dedicated rf synthesis system. The scheme allows us to transmit the laser frequency stability to a microwave oscillator without need to impress it on the OFC spectrum. In fact, the generated frequency is immune to fluctuations of the OFC spectrum [25,26]. With these features, the transfer oscillator scheme offers significant advantages over direct optical-to-microwave conversion schemes that rely on phase locking an OFC to a stable laser [19,21,22,24].

The optical frequency reference used in our experiment is a laser with sub-hertz linewidth that is stabilized to the electric-quadrupole $^2S_{1/2}(F = 0)$ - $^2D_{3/2}(F' = 2)$ transition of a single trapped $^{171}$Yb$^+$ ion at the frequency $\nu(\text{Yb}^+) \approx 688$ THz. Using the OFC-stabilized interrogation oscillator with the fountain clock CSF2 of PTB (Physikalisch-Technische Bundesanstalt) [27,28], we demonstrate a fountain stability of $\sigma_y(\tau) = 4.1 \cdot 10^{-14} (\tau/\text{s})^{-1/2}$ for operation with a laser-cooled atomic beam source.

The utility for absolute frequency measurements is demonstrated through a new measurement of the frequency $\nu(\text{Yb}^+)$, which is recognized as a secondary representation of the SI second [29]. Here CSF2 is operated with the OFC-stabilized interrogation oscillator and simultaneously $\nu(\text{Yb}^+)$ is measured with the fountain clock CSF1 operating with a conventional quartz-based microwave synthesis system [30,31]. In an averaging time of 65 h, $\nu(\text{Yb}^+)$ is determined with a total relative uncertainty of $5.2 \cdot 10^{-16}$, which corresponds to an improvement by a factor of two relative to the previous measurement [32]. Due to a substantially reduced systematic uncertainty of the $\text{Yb}^+$ single-ion frequency standard, the uncertainty of the measured value of $\nu(\text{Yb}^+)$ is now essentially determined by the systematic uncertainty of the fountain references.

## II. EXPERIMENTAL SETUP

### A. $^{171}\text{Yb}^+$ optical frequency standard

A single $^{171}\text{Yb}^+$ ion is confined in a rf Paul trap and laser cooled by excitation of the 370-nm $^2S_{1/2}$ - $^2P_{1/2}$ resonance transition and of a repumping transition at 935 nm [6,32]. Each measurement cycle consists of a period of state detection, laser cooling, and preparation in the $F = 0$ ground-state sublevel ($\approx$ 35 ms) and a period where the cooling and repumping lasers are blocked by mechanical shutters and a 30-ms rectangular interrogation pulse resonant with the $^2S_{1/2}(F = 0)$ - $^2D_{3/2}(F' = 2, m_F = 0)$ transition at 688 THz (436 nm) is applied. The trap is magnetically shielded by a mu-metal enclosure. During excitation of the reference transition, a set of coils inside the shielding allows us to set the magnetic field at the trap center to three orientations that are mutually orthogonal with an uncertainty of $\leq 1°$, which enables an efficient cancellation of tensorial shifts [33].

The $\text{Yb}^+$ reference transition is interrogated by the frequency-doubled output of an extended-cavity diode laser operating at 871 nm (laser frequency $\nu_L \approx 344$ THz). By locking to an environmentally isolated high-finesse cavity, we obtain a laser linewidth of less than 1 Hz and a relative frequency instability of $\sigma_y(\tau) \leq 2 \cdot 10^{-15}$ for 0.3 s $\leq \tau \leq$ 30 s [34]. With probe pulses of 30 ms duration, the reference transition is resolved with a Fourier-limited linewidth of $\approx$ 26 Hz FWHM. The resonant excitation probability of $\approx$ 70% is essentially limited by the natural lifetime of the $^2D_{3/2}$ level of 52 ms.

The frequency of the 688-THz probe light is tuned by two acousto-optic modulators that introduce frequency shifts $2\nu_L + \nu_{AOM1}$ and $\nu_{AOM2}$ (see left part of Fig. 1). $\nu_{AOM1}$ is alternately stepped up and down by 13 Hz in subsequent cycles in order to obtain an error signal for the servo system that steers $2\nu_L + \nu_{AOM1}$ to the line center of the Yb$^+$ reference transition by means of AOM2. The error signal is processed by a second-order integrating servo algorithm that minimizes slowly varying offsets due to drift of the high-finesse cavity [35]. In order to monitor the magnitude of tensorial shifts of the transition frequency, the direction of the applied magnetic field is periodically switched between two of three perpendicular orientations after four cycles. For one orientation, $\nu_{AOM2}$ is controlled by the servo system while $\nu_{AOM1}$ is referenced by a maser signal. The frequency offset that appears in the other orientation is balanced by controlling $\nu_{AOM1}$ so that the stability of $\nu_L$ is not affected by the alternating servo operation. The time constant of the lock of $\nu_L$ to the Yb$^+$ transition is in the range of 10 s. At longer times, the instability of $\nu_L$ is essentially determined by quantum projection noise and is expected to be near $\sigma_y(\tau) = 1 \cdot 10^{-14} (\tau/\text{s})^{-1/2}$ [35,36].

B. Caesium fountain clock

The setup and the general operating conditions of the caesium fountain CSF2 are described in Refs. [27,28]. In the standard operation mode of CSF2, Cs atoms are collected from thermal background vapor in an optical molasses using polarization gradient cooling. The collisional shift of the hyperfine transition frequency is determined through an adiabatic passage technique that reduces the density of launched Cs clouds by exactly a factor of two [37,38]. As an alternative to loading from thermal vapor, a considerably larger number of atoms can be loaded into the molasses more quickly from a source of low-velocity atoms extracted from a magneto-optical trap [39,40].

As shown on the right of Fig. 1, the interrogation signal of CSF2 is produced by a commercial dielectric resonator microwave oscillator (DRO) [41]. The DRO output frequency $\nu_{DRO} \approx 9.6$ GHz is stabilized to the optical frequency reference by a phase locked loop that involves the synthesized transfer frequency $\nu_T$. As described below in Sec. II C, the frequency $\nu_T$ displays the phase fluctuations of the DRO relative to the Yb$^+$ interrogation laser. The servo loop that controls $\nu_{DRO}$ has a bandwidth of approximately 50 kHz and an integrating servo characteristic at Fourier frequencies $f \leq 12$ kHz to enhance the suppression of phase noise close to the carrier.

The DRO output frequency is shifted to the range of the Cs hyperfine transition by a divider system that is controlled with a digital resolution of $2 \cdot 10^{-17}$. The Cs atoms are excited alternately on both sides of the central fringe of the Ramsey resonance in order to generate an error signal. The error signal adjusts the frequency division ratio so that the mean excitation frequency is stabilized at resonance center. By taking into account the systematic biases of the Cs resonance frequency, $\nu_{DRO}$ is determined with reference to the unperturbed frequency $\nu_{Cs}$ [27,28].

C. Optical-to-microwave conversion

The experiment uses a commercial OFC based on an Er$^{3+}$-doped fiber laser that emits in the wavelength range around 1550 nm with a pulse repetition rate $\nu_{rep} \approx 250$ MHz. The OFC includes elements for generating the carrier offset frequency $\nu_{CEO}$ and for producing output at 1740 nm that is frequency doubled to the range of $\nu_L$ (see Fig. 1). The frequencies $\nu_{rep}$ and $\nu_{CEO}$ are stabilized with low bandwidth ($\approx 300$ Hz) to a maser reference to reduce thermal drifts. The fluctuations of the comb line frequencies near $\nu_L$ are typically in the range of 100 kHz with Fourier frequencies up to 50 kHz.

As shown in Fig. 1, the transfer frequency $\nu_T$ is synthesized from the input frequencies $\nu_{DRO}$, $\nu_{CEO}$, $k \cdot \nu_{rep}$, and $\nu_x$. Here $k = 38$ denotes the order of the harmonic of $\nu_{rep}$ nearest to $\nu_{DRO}$ and $\nu_x$ is the frequency of the beat between $\nu_L$ and the nearest mode of the frequency-doubled OFC spectrum. With $m$ denoting the order number of this mode, $\nu_x$ is given by

$$\nu_x = \nu_L - m \cdot \nu_{rep} - 2 \cdot \nu_{CEO} \quad . \tag{2}$$

The prefactor 2 of $\nu_{CEO}$ in Eq. (2) takes into account that $\nu_x$ is produced by the frequency-doubled OFC output. In a setup comprising passive and tracking rf bandpass filters, rf mixers, a phase locked harmonic oscillator, and digital frequency dividers, we generate the intermediate frequency

$$\nu_1 = l \cdot (\nu_{DRO} - k \cdot \nu_{rep}) \quad . \tag{3}$$

It will become clear below that the factor $l$ determines the sensitivity with which variations of $\nu_{DRO}$ are tracked. In our setup, $l = 8$ because bandwidth requirements would make it difficult to increase it further. We also generate

$$\nu_2 = \frac{1}{c} \cdot \left( \frac{1}{2} \nu_x + \nu_{CEO} \right) \quad , \tag{4}$$

where the divisor $c$ is a preset rational number. Inserting Eq. (2) into Eq. (4) shows that $\nu_2$ is independent of $\nu_{CEO}$. For $c = m/(2kl)$, the transfer frequency $\nu_T = \nu_2 - \nu_1$ is determined alone by $\nu_L$ and $\nu_{DRO}$:

$$\nu_T = \frac{1}{2c}\nu_L - l \cdot \nu_{DRO} \quad . \tag{5}$$

We realize the frequency division by $c = m/(2kl) \approx 2242.1$ with three low-noise 8:1 dividers and a direct digital synthesis (DDS) stage. Between the divider stages, the frequency is shifted up by mixing with a 100-MHz maser signal in order to enable operation in the range of 20...200 MHz where the noise level of the employed rf components is minimal. In the DDS, the exact value of $c$ is approximated with a maximum error $\delta c/c = 2^{-48} \approx 4 \cdot 10^{-15}$. This leads to a crosstalk from fluctuations of $\nu_{rep}$ to $\nu_T$ of relative order $2kl\delta c/c$ which is negligible under our conditions.

Since its synthesis is phase coherent, $\nu_T$ can be regarded as the beat frequency of the two virtual frequencies $\nu_L/(2c)$ and $l \cdot \nu_{DRO}$ [20,25,26]. Frequency and phase fluctuations of the DRO appear magnified by a factor $l = 8$ in $\nu_T$ so that they can be detected and controlled with particularly high signal-to-noise ratio. The phase noise spectrum of the laser-stabilized DRO is shown in Fig. 2(a). This measurement was performed using a second OFC system independently referenced to the same laser. For comparison, Fig. 2 also shows the specified phase noise performance of the unstabilized DRO and the noise floor of a recently developed cryocooled sapphire oscillator [17]. It appears that the phase noise level of the stabilized DRO is comparable to that of the cryogenic oscillator. The increase at Fourier frequencies $\leq 10$ Hz is mainly caused by the limited selectivity of the rf spectrum analyzer used for the measurement. Significant noise peaks appear only at the line frequency (50 Hz) and its harmonics. Microwave signals generated by cryocooled sapphire oscillators and laser-locked OFC systems typically show additional phase noise peaks caused by vibrations and acoustic perturbations. In our case, such features are well suppressed because fluctuations of $\nu_{rep}$ and $\nu_{CEO}$ are not passed to $\nu_T$.

According to Eqs. (2) and (3), the relation between the optical frequency $\nu_L$ and $\nu_{DRO}$ is given by

$$\nu_L - \frac{m}{k}\nu_{DRO} = \nu_x + 2 \cdot \nu_{CEO} - \frac{m}{k \cdot l}\nu_1 \quad . \tag{6}$$

The result is independent of $\nu_{rep}$ and $\nu_{CEO}$ because the respective contributions cancel out on the right hand side of Eq. (6). In our experiment, $\nu_{DRO}$ is directly measured by CSF2 as described in Sec. II B, and $\nu_x$, $\nu_{CEO}$, and $\nu_1$ are registered by three counters referenced to $\nu_{Cs}$ through a maser signal. The counters are read out synchronously at a rate of 1 s$^{-1}$ to calculate $\nu_L - (m/k)\nu_{DRO}$.

## III. RESULTS

A. Fountain clock stability

For frequency measurements, CSF2 is alternately operated with high and low Cs densities to monitor the collisional shift in real time. The loading time and the duty cycle of the two density modes are chosen such that for a given oscillator phase noise level the total uncertainty is minimized. [23,27]. The stability of CSF2 in the high-density mode relative to a hydrogen maser is shown in Fig. 3(a). Here the fountain was operated with a quartz-controlled interrogation oscillator [27]. For averaging times $\tau > 3000$ s, the observed instability is dominated by the drift of the employed maser reference. The inferred fountain stability is $\sigma_y(\tau) = 1.9 \cdot 10^{-13} (\tau/\text{s})^{-1/2}$. Even though higher detected atom numbers are achievable with longer loading time, the stability does not improve because the local oscillator noise becomes the dominant noise source in the resonance signal [14].

If CSF2 is operated with the Yb$^+$-stabilized interrogation oscillator setup under otherwise unchanged conditions, the fountain stability is improved. Increasing the detected atom number by increasing the atom loading time compared to Fig. 3(a) results in a further improved stability depicted in Fig. 3(b). The data show a fountain stability of $\sigma_y(\tau) = 1.0 \cdot 10^{-13} (\tau/\text{s})^{-1/2}$ in the high-density operation mode. We find that the SNR of the fountain error signal is proportional to the square root of the detected atom number signal. The same SNR is obtained if the ground-state hyperfine transition is excited with two resonant $\pi/4$ pulses which minimizes the influence of interrogation oscillator noise. These observations confirm that the fountain stability is limited by quantum projection noise without stability degradation by the interrogation oscillator [23]. Under these conditions the number of detected Cs atoms can be inferred directly from the observed SNR. For Fig. 3(b), the calculated detected atom number is $1.1 \cdot 10^5$.

The data of Fig. 3(c) demonstrate the fountain stability in high-density mode of operation if the optically stabilized interrogation oscillator is used together with the molasses loaded from a low-velocity atomic beam. The stability improvement over Fig. 3(b) is the result of a larger number of detected atoms and a reduced cycle time. The observed stability of $\sigma_y(\tau) = 4.1 \cdot 10^{-14} (\tau/\text{s})^{-1/2}$ is comparable to that obtained in an absolute frequency measurement performed with a cryogenic sapphire fountain interrogation oscillator [9]. Since the uncertainty evaluation of CSF2 is still incomplete for this operation mode, it was not adopted for the measurement of $\nu(\text{Yb}^+)$.

B. Systematic uncertainty of the $^{171}\text{Yb}^+$ frequency standard

The leading systematic shifts of the $^{171}\text{Yb}^+$ frequency standard and the associated uncertainty contributions are listed in Table I. The largest shift is the second-order Zeeman shift due to the static magnetic field that defines the quantization axis during excitation of the Yb$^+$ reference transition. The calculation of this shift and its uncertainty makes use of a new measurement of the hyperfine splitting frequency in the $^2D_{3/2}$ state and is described in the Appendix. For the measurement of $\nu(\text{Yb}^+)$, the magnetic flux density was chosen to obtain a Zeeman splitting frequency of $\nu_Z = 30.0$ kHz between adjacent sublevels of the $^2D_{3/2}(F' = 2)$ state. For the three applied field orientations, $\nu_Z$ was set with an uncertainty of 0.1 kHz by excitation of the $m_F = 0$ - $m_F = 0$ and $m_F = 0$ - $m_F = \pm 2$ components of the $^2S_{1/2}(F = 0)$ - $^2D_{3/2}(F' = 2)$ transition. The drift of $\nu_Z$ over the $\nu(\text{Yb}^+)$ measurement period was found to be smaller than the setting uncertainty. As shown in the Appendix, the uncertainty of the calculated second-order Zeeman shift of 0.666(4) Hz is determined rather by the uncertainty of $\nu_Z$ than by the uncertainty of the atomic parameters used in the calculation.

The quadrupole shift of the $^2D_{3/2}$ ($F' = 2$, $m_F = 0$) level due to electric stray field gradients and the shift arising from the tensorial quadratic Stark effect are cancelled by averaging over three mutually perpendicular orientations of the quantization axis [33]. In the course of the absolute frequency measurement, we determined the Yb$^+$ transition frequency sequentially for an orthogonal set of magnetic field orientations. At the same time the frequency differences between three pairs of field directions were registered using the alternating servo scheme described in Sec. II A. From the latter the tensorial shifts for the three field directions were calculated as 0.41(2) Hz, -0.47(2) Hz, and 0.06(2) Hz where the uncertainty is predominantly statistical. The shifts remained stable within uncertainty during the measurement period. The evaluated absolute frequency $\nu(\text{Yb}^+)$ is the average of the transition frequencies measured in the three field directions. Taking into account the observed magnitude of the shifts and the orientation accuracy of the applied magnetic field, we infer an uncertainty contribution of 0.01 Hz due to imperfect cancellation of the quadrupole and tensorial Stark shift.

The motion of the trapped ion at the trap drive frequency was minimized by stray-field compensation voltages that were adjusted as described in Ref. [36]. The $\nu(\text{Yb}^+)$ measurement reported here was conducted several weeks after loading the ion. Therefore initial variations of the stray field had decayed and it remained stable within the compensation uncertainty. By optical spectroscopy of the secular oscillation sidebands, the amplitude of the residual thermal motion was found to be close to the Doppler limit. Under these conditions the second-order Doppler shift and the second-order scalar Stark shift due to ion motion amount to $3 \cdot 10^{-3}$ Hz and $9 \cdot 10^{-3}$ Hz, and we assume uncertainty contributions of half the shift. Changes of the drift rate of the cavity stabilizing the probe laser frequency had a magnitude below $10^{-3}$ Hz$/s^2$ so that in our case the maximum servo error in the stabilization to the center of the Yb$^+$ reference transition is 0.025 Hz [5].

The dominant contribution to the systematic uncertainty of $\nu(\text{Yb}^+)$ arises from the second-order Stark shift induced by the ambient blackbody radiation. Earlier measurements of the static scalar differential polarizability indicate a shift of -0.36(7) Hz at temperature $T = 300$ K [36,42]. It appears that so far no atomic structure calculations have been presented for $\text{Yb}^+$ that determine the differential polarizability and the blackbody shift of the $^2S_{1/2}$ - $^2D_{3/2}$ transition with lower uncertainty. A variation $\delta T$ of the effective temperature of the thermal radiation acting on the ion changes the blackbody shift by $\approx$ -5·10$^{-3}$ Hz·$(\delta T / \text{K})$. An investigation of blackbody radiation sources indicates an increase $\delta T = 2(1)$ K above ambient temperature for our system [43]. The resulting change and additional uncertainty of the blackbody shift is presently negligible compared to the contribution of the polarizability uncertainty.

As summarized in Table I, we realize $\nu(\text{Yb}^+)$ with a fractional systematic uncertainty of 1.1·10$^{-16}$ which is more than a factor of four lower than in the previous measurement [32].

C. Absolute frequency measurement

We determined the unperturbed frequency $\nu(\text{Yb}^+)$ of the $^{171}\text{Yb}^+$ $^2S_{1/2}(F = 0)$ - $^2D_{3/2}(F' = 2)$ transition from data comprising a total measurement time of 64.5 h within a 5-day period in 8/2012. The datasets yield time averages of $\nu_{\text{DRO}}$ measured by CSF2 and of $\nu_{\text{L}} - (m/k)\nu_{\text{DRO}}$ with $\nu_{\text{L}}$ locked to the $^{171}\text{Yb}^+$ reference transition. The instability of this measurement is dominated by quantum projection noise of the fountain error signal and amounts to $\sigma_y(\tau) = 1.4 \cdot 10^{-13} (\tau/\text{s})^{-1/2}$. It is increased by a factor of 1.36 relative to the high-density operation of CSF2 shown in Fig. 3(b) because the fountain is operated alternately with high and low atom density to determine the collisional shift. The uncertainty of the collisional shift is the leading contribution to the systematic uncertainty of CSF2 and amounts to 5.5·10$^{-16}$.

The frequency $\nu(\mathrm{Yb}^+)$ was simultaneously measured by the fountain CSF1. Here the frequency difference between the CSF1 quartz oscillator and a hydrogen maser was determined by a phase comparator while the maser frequency was linked to the probe laser frequency $\nu_L$ by means of the OFC. The CSF1 atom cloud density was not changed during the $\nu(\mathrm{Yb}^+)$ evaluation and the collisional shift and its uncertainty were evaluated in a separate measurement. The systematic uncertainties of CSF1 and CSF2 are $7.4 \cdot 10^{-16}$ and $5.9 \cdot 10^{-16}$ and the statistical uncertainties of the measurements are $3.2 \cdot 10^{-16}$ and $2.8 \cdot 10^{-16}$. The fractional difference between the frequencies measured by CSF1 and CSF2 is $2.6 \cdot 10^{-16}$. The weighted mean yields the absolute frequency of the $^{171}\mathrm{Yb}^+$ $^2S_{1/2}(F = 0)$ - $^2D_{3/2}(F' = 2)$ transition as $\nu(\mathrm{Yb}^+) = 688\,358\,979\,309\,307.82(36)$ Hz at $T = 0$ K. The total uncertainty of 0.36 Hz (relative uncertainty $5.2 \cdot 10^{-16}$) is dominated by the systematic uncertainty of the caesium fountain references. Figure 4 shows that the present result is in good agreement with those obtained in the last seven years [32,44,45] with different experimental configurations and systematic shift evaluation methods.

## IV. SUMMARY AND OUTLOOK

We have described and characterized a new experimental configuration for absolute measurements of optical clocks by means of an optical frequency comb generator (OFC) and a caesium fountain. Besides the function to measure the optical frequency, the OFC also transfers the stability of the optical standard to the microwave range to enhance the stability of the fountain reference. We have utilized this scheme to measure the absolute frequency of the 688-THz $^2S_{1/2}(F = 0)$ - $^2D_{3/2}(F = 2)$ electric-quadrupole transition of a single trapped $^{171}\mathrm{Yb}^+$ ion. The achieved fractional frequency uncertainty of $5.2 \cdot 10^{-16}$ is essentially limited by the systematic uncertainty of the employed Cs fountain clocks and is one of the most precise determinations of an optical transition frequency.

In our experiment we use the transfer oscillator scheme to transmit the stability of a laser frequency to a microwave oscillator without imprinting it on the OFC spectrum [25,26]. One advantage of this scheme is that the transfer fidelity is not limited by the bandwidth and accuracy with which the OFC spectrum is controlled. Another advantage is that one OFC can serve various tasks without interference. For example, frequency ratio measurements between three or more optical and microwave standards can be performed simultaneously, thereby enabling stringent consistency tests.

Among the atoms and ions considered so far for optical frequency standards, $Yb^+$ is unique because it provides two narrow-linewidth transitions from the ground state that are attractive reference transitions: the $^2S_{1/2}$ - $^2D_{3/2}$ transition considered in this paper and the highly forbidden electric-octupole transition $^2S_{1/2}$ - $^2F_{7/2}$ at 642 THz [6,46,47]. The physical characteristics of both transitions and in particular the sensitivities of the transition frequencies to external fields differ substantially [6,42,48,49]. The high degree of control of frequency shift effects that was established in the present frequency measurement is instrumental also in the future development of the $^{171}Yb^+$ $^2S_{1/2}$ - $^2F_{7/2}$ standard. We finally note that our measurements of the $^{171}Yb^+$ $^2S_{1/2}$ - $^2D_{3/2}$ and $^2S_{1/2}$ - $^2F_{7/2}$ transition frequencies yield the ratio of the corresponding energy level intervals with a relative uncertainty below $1 \cdot 10^{-15}$. We do not know of another intrinsic atomic or molecular level interval ratio that is known with comparable relative accuracy.

## ACKNOWLEDGEMENTS

We thank C. Sanner, P. Balling and M. Dolezal for sharing their results on investigations of blackbody radiation in the employed ion trap. This work was supported by the European Metrology Research Programme (EMRP) in project SIB04. The EMRP is jointly funded by the EMRP participating countries within EURAMET and the European Union. NH acknowledges funding from Graduiertenkolleg 1729 of DFG.

**APPENDIX: CALCULATION OF THE SECOND-ORDER ZEEMAN SHIFT**

The second-order Zeeman shift $\Delta_{Z2}$ of the $^2S_{1/2}(F = 0)$ - $^2D_{3/2}(F' = 2, m_F = 0)$ transition frequency of $^{171}$Yb$^+$ due to an applied magnetic flux density $B$ is calculated as the difference of the shifts of the upper and lower level, quantified by the corresponding shift coefficients $\alpha_{S0}$ and $\alpha_{D2}$:

$$\Delta_{Z2} = (\alpha_{D2} - \alpha_{S0}) B^2 \quad . \tag{A1}$$

The coefficients $\alpha_{S0}$ and $\alpha_{D2}$ can be calculated from second-order perturbation theory in the state basis $\langle IJFm_F |$ with the quantum numbers $I$, $J$, $F$, and $m_F$ denoting the nuclear spin, the electronic angular momentum, the total angular momentum, and its projection on the direction of $B$. Neglecting the contribution of levels outside the respective hyperfine manifolds, in our case $\alpha_{S0}$ and $\alpha_{D2}$ are given by:

$$\alpha_{S0} = -\frac{\mu_B^2 \left(g_{\frac{1}{2}} + g_I'\right)^2}{h^2} \cdot \frac{|\langle \frac{1}{2}\frac{1}{2} 00 | J_z | \frac{1}{2}\frac{1}{2} 10 \rangle|^2}{\Delta_S} \tag{A2a}$$

and

$$\alpha_{D2} = \frac{\mu_B^2 \left(g_{\frac{3}{2}} + g_I'\right)^2}{h^2} \cdot \frac{|\langle \frac{1}{2}\frac{3}{2} 10 | J_z | \frac{1}{2}\frac{3}{2} 20 \rangle|^2}{\Delta_D} \quad . \tag{A2b}$$

The terms containing the matrix elements of the $z$-component of the angular momentum operator $J_Z$ evaluate to [50]

$$|\langle \tfrac{1}{2}\tfrac{1}{2} 00 | J_z | \tfrac{1}{2}\tfrac{1}{2} 10 \rangle|^2 = |\langle \tfrac{1}{2}\tfrac{3}{2} 10 | J_z | \tfrac{1}{2}\tfrac{3}{2} 20 \rangle|^2 = 1/4 \quad . \tag{A2c}$$

In Eq. (A2a) and (A2b), $h$ denotes the Planck constant, $\mu_B$ is the Bohr magneton, $g_{\frac{1}{2}}$ ($g_{\frac{3}{2}}$) is the electron $g_J$-factor in the $^2S_{1/2}$ ($^2D_{3/2}$) state of Yb$^+$, and $g_I' = \mu_I / (\mu_B I)$ is the nuclear $g$-factor associated with the nuclear magnetic moment $\mu_I$. $\Delta_S$ ($\Delta_D$) denotes the hyperfine splitting frequency in the $^2S_{1/2}$ ($^2D_{3/2}$) state. Eq. (A2a) can also be deduced from the low-field limit of the Breit-Rabi formula [51].

In previous calculations of $\Delta_{Z2}$, a significant uncertainty contribution arose from the lack of precise information on the hyperfine splitting frequency $\Delta_D$. Whereas $\Delta_S$ is known with high accuracy [52], $\Delta_D$ has been known only with an uncertainty of $\approx 2.5$ % ($\Delta_D = 0.86(2)$ GHz according to Ref. 53). In order to gain more accurate information, we modified our experimental setup to remeasure $\Delta_D$ using a laser-rf double-resonance spectroscopy scheme. In an extension of the cycle sequence described in Sec. II A, the $^2D_{3/2}(F' = 2, m_F = 0)$ is populated by a resonant laser pulse and subsequently a 30-ms rf pulse is applied. Resonant excitation of the $^2D_{3/2}(F' = 2 - F' = 1)$ transition by the rf field is detected by the reduced probability to find the ion in the $^2D_{3/2}(F' = 2)$ state in the next detection period. In this way we determine the $^2D_{3/2}(F' = 1 - F' = 2)$ interval as $\Delta_D = 857.178078(5)$ MHz.

Experimental values of $g_{1/2} = 1.998$ and $g_{3/2} = 0.802$ can be found in tabulated spectroscopic data [54]. We estimate that the uncertainty of those data is in the range of $3 \cdot 10^{-3}$. A recent calculation of g-factor anomalies in a number of atomic systems yields $g_{1/2} = 2.0031(3)$ and $g_{3/2} = 0.7992(1)$ for Yb$^+$ where the uncertainty is the estimated maximum error of the calculation [55]. With these more accurate $g_J$-factor values and $g'_I = +5.38 \cdot 10^{-4}$ [56], the shift coefficients calculated by Eq. (A2) are $\alpha_{S0} = -15.551(5)$ mHz$/(\mu T)^2$ and $\alpha_{D2} = 36.545(11)$ mHz$/(\mu T)^2$ so that $\Delta_{Z2} = 52.096(16)$ mHz$/(\mu T)^2$.

The Zeeman frequency is given by $\nu_Z = g_F \mu_B B/h$ with $g_F = \frac{3}{4} g_{3/2} - \frac{1}{4} g'_I = 0.59930(9)$ for the $^2D_{3/2}(F' = 2)$ state [51]. The value set in the measurement of $\nu(Yb^+)$, $\nu_Z = 30.0(1)$ kHz, corresponds to an applied flux density $B = 3.58(1) \mu T$ and leads to a second-order Zeeman shift of the $^2S_{1/2}(F = 0)$ - $^2D_{3/2}(F' = 2, m_F = 0)$ transition frequency of $\Delta_{Z2} = 0.666(4)$ Hz.

**Table I**

TABLE I. Leading systematic shifts $\delta\nu/\nu(\text{Yb}^+)$ and associated uncertainty contributions $u_B/\nu(\text{Yb}^+)$ of the 688-THz $^{171}\text{Yb}^+$ single-ion optical frequency standard.

| Effect | $\delta\nu/\nu(\text{Yb}^+)$ ($10^{-18}$) | $u_B/\nu(\text{Yb}^+)$ ($10^{-18}$) |
|---|---|---|
| Blackbody radiation shift | -524 | 102 |
| Quadrupole shift and tensorial second-order dc Stark shift | 0 | 14 |
| Scalar second-order dc Stark shift | -13 | 7 |
| Second-order Zeeman shift | 968 | 7 |
| Second-order Doppler shift | -5 | 3 |
| Servo error | 0 | 36 |
| Total | 426 | 110 |

**Figure 1**

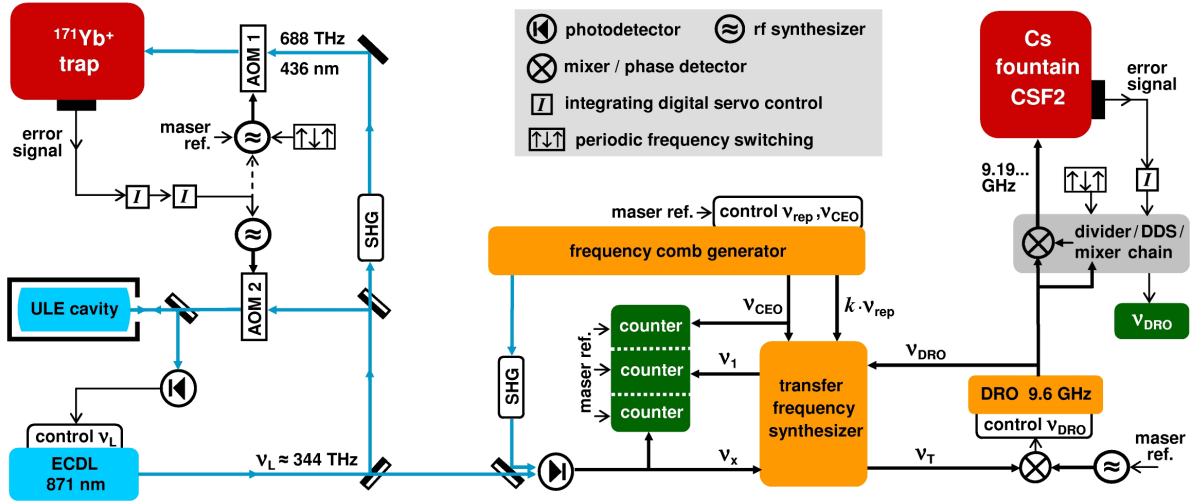

FIG. 1: Schematic of the experimental setup, showing components of the Yb$^+$ optical frequency standard (left), of the frequency comb generator system (center), and of the Cs fountain reference (right). In the realized setup, output from the 871 nm extended-cavity diode laser (ECDL) is forwarded to the frequency comb generator by ≈5 m polarization-maintaining fiber.

**Figure 2**

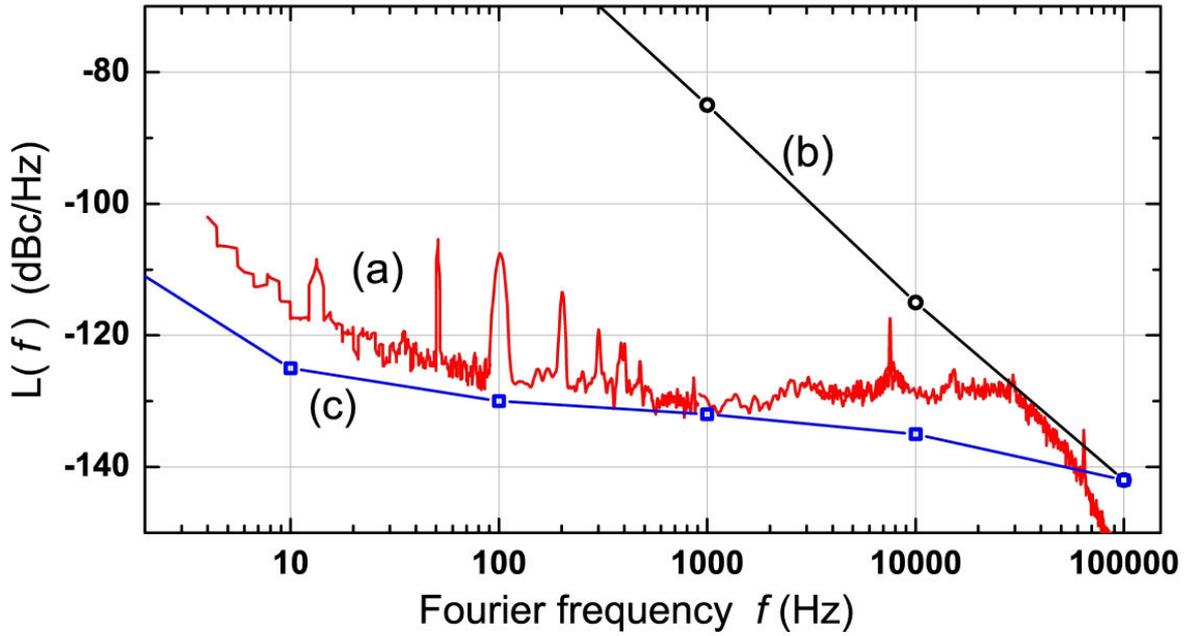

FIG. 2: Single-sideband phase noise power density $L$ relative to the carrier as a function of offset (Fourier) frequency $f$. (a), red: measured phase noise spectrum of the employed dielectric resonator oscillator (DRO) for the case that its frequency is locked to the Yb$^+$ interrogation laser using the transfer oscillator scheme (see text); (b), black: specified phase noise of the unstabilized DRO; (c), blue: noise floor of the cryocooled sapphire microwave oscillator described in Ref. [17].

**Figure 3**

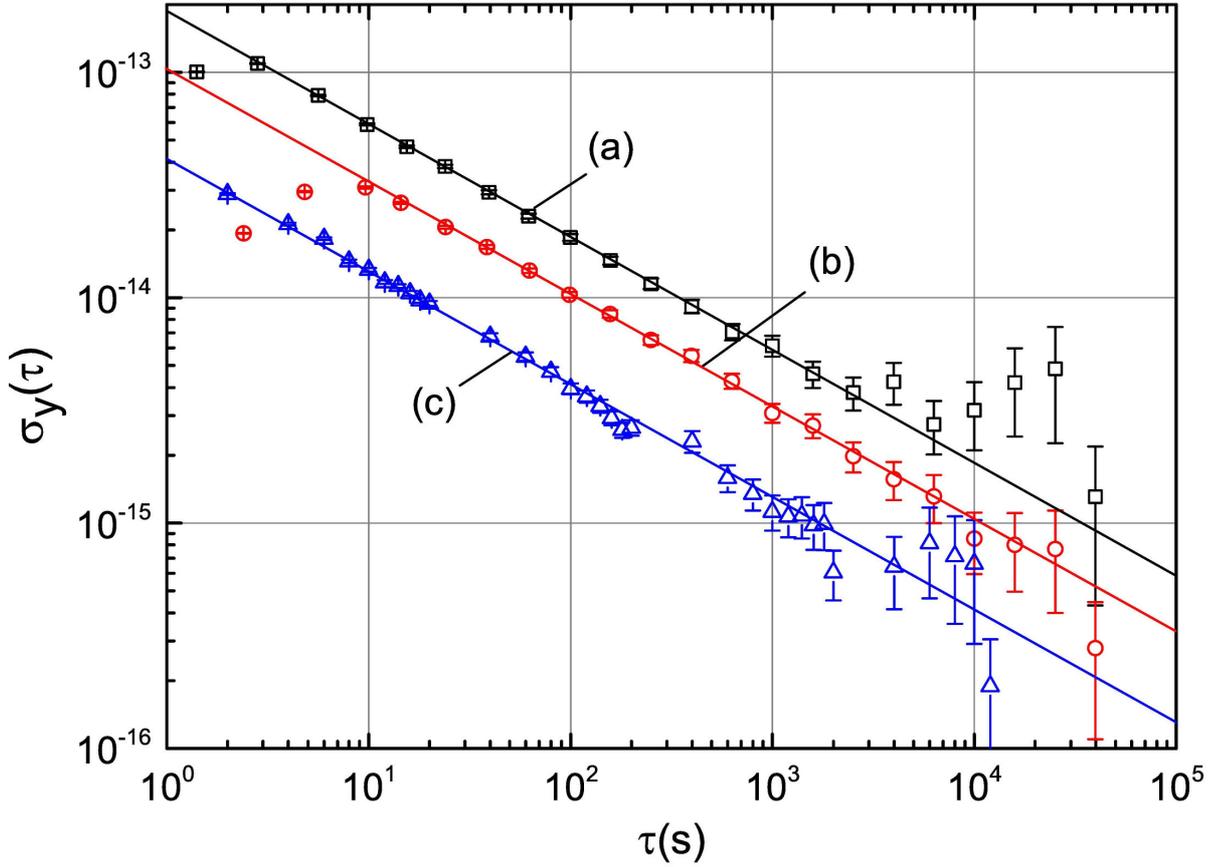

FIG. 3: Measured Allan standard deviations characterizing the stability of CSF2. (a), black: quartz-based interrogation oscillator, maser reference, molasses loading time 0.5 s. (b), red: interrogation by a dielectric resonator oscillator (DRO) stabilized by a laser locked to the $^2S_{1/2}$ - $^2D_{3/2}$ transition of a trapped Yb$^+$ ion (see text), molasses loading time 2 s. (c), blue: interrogation oscillator as in (b) combined with molasses loading from a low-velocity atomic beam, loading time 0.8 s. The lines denote the inferred instabilities assuming a dependence according to $\sigma_y(\tau) \propto \tau^{-1/2}$.

**Figure 4**

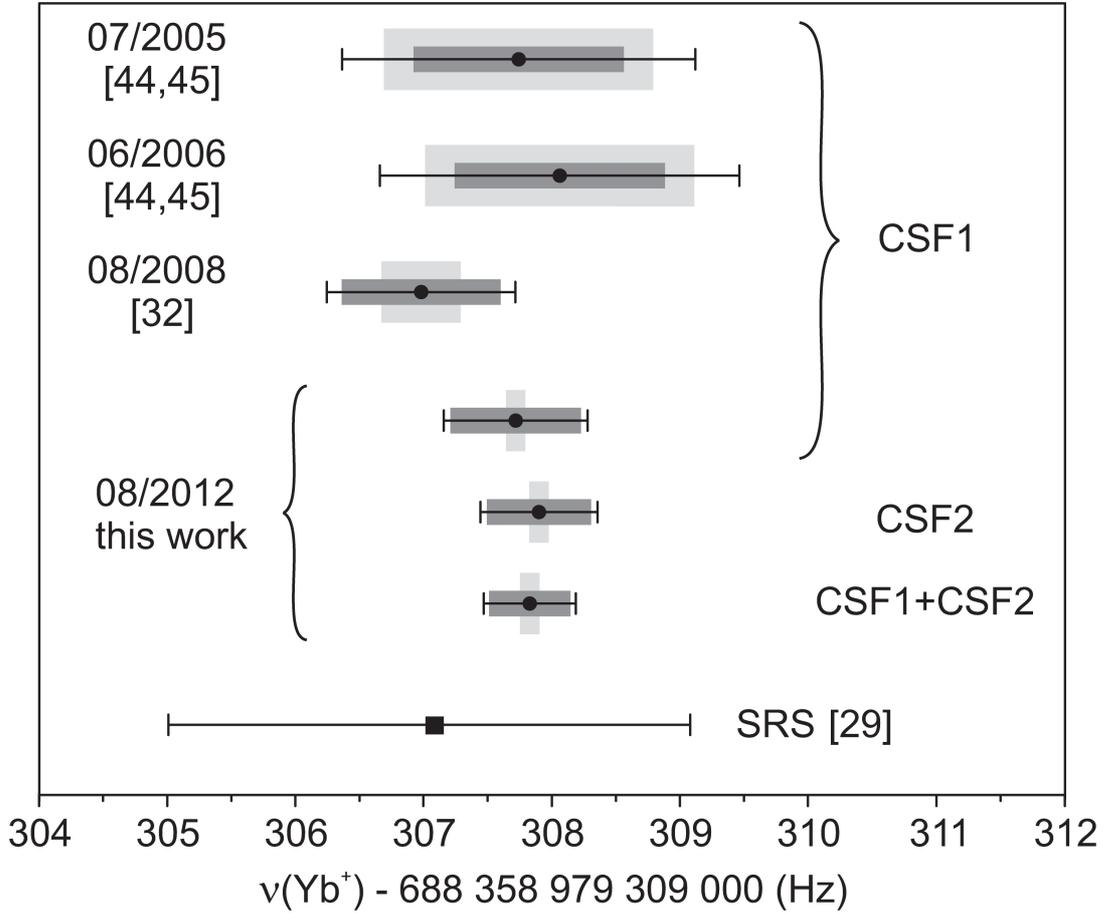

FIG. 4: Results and uncertainties of the present and of previous measurements of the absolute frequency $\nu(Yb^+)$ of the $^{171}Yb^+$ $^2S_{1/2}(F=0)$ - $^2D_{3/2}(F'=2)$ transition. The corresponding References are indicated in the Figure. Unlike the previously published data, the frequencies given here contain a blackbody shift correction of 0.36 Hz as applied in the present measurement. The black error bars show the total measurement uncertainty including the statistical contribution. The widths of the dark and light grey rectangles show the systematic uncertainties of the Cs reference and of the $Yb^+$ frequency standard. The data point "SRS" shows the recommended frequency and uncertainty of $\nu(Yb^+)$ as a secondary representation of the SI second.